\documentclass[aip,jcp,reprint]{revtex4-2}

\usepackage{amsmath,amssymb,graphicx}
\usepackage{dcolumn}
\usepackage{bm}
\usepackage[colorlinks=true,linkcolor=cyan,citecolor=cyan,urlcolor=black]{hyperref}

\begin{document}

\bibliographystyle{aipnum4-2}  

\title{Thermodynamic Stability and Kinetic Control of Capsid Morphologies in Hepatitis B Virus}

\author{Kevin Yang}
\affiliation{Biophysics Graduate Program, University of California, Riverside, California 92521, USA}

\author{Juana Martin Gonzalez}
\affiliation{Department of Physics and Astronomy, University of California, Riverside, California 92521, USA}

\author{Alireza Ramezani}
\affiliation{Department of Physics and Astronomy, University of California, Riverside, California 92521, USA}

\author{Paul van der Schoot}
\affiliation{Department of Applied Physics and Science Education, Eindhoven University of Technology, Postbus 513, 5600 MB Eindhoven, The Netherlands}

\author{Roya Zandi}
\email{royaz@ucr.edu}
\affiliation{Biophysics Graduate Program, University of California, Riverside, California 92521, USA}
\affiliation{Department of Physics and Astronomy, University of California, Riverside, California 92521, USA}

\date{\today}

\begin{abstract}
Polymorphism has been observed in viral capsid assembly, demonstrating the ability of identical protein dimers to adopt multiple geometries under the same solution conditions. A well-studied example is the hepatitis B virus (HBV), which forms two stable capsid morphologies both \textit{in vivo} and \textit{in vitro}. These capsids differ in diameter, containing either 90 or 120 protein dimers. Experiments have shown that their relative prevalence depends on the ionic conditions of the solution during assembly. We developed a model that incorporates salt effects by altering the intermolecular binding free energy between capsid proteins, thereby shifting the relative thermodynamic stability of the two morphologies. This model reproduces experimental results on the prevalence ratios of the large and small HBV capsids. We also constructed a kinetic model that captures the time-dependent ratio of the two morphologies under subcritical capsid concentrations, consistent with experimental data.
\end{abstract}

\maketitle


%
%

\section{Introduction}\phantomsection\label{sec:introduction}
The majority of viruses consist of a protein shell, or capsid, that packages and protects their genetic material. In most spherical viruses, the capsid adopts an icosahedral architecture, which can be classified by a triangulation number ($T$). This structural index determines the number of protein subunits in the shell, calculated as $60 \times T$, where $T = h^2 + hk + k^2$ and $h, k$ are non-negative integers. While capsids reliably assemble into a specific structure under many \textit{in vivo} and \textit{in vitro} conditions, in some cases polymorphism is observed, in which capsid proteins spontaneously assemble into particles of varying sizes~\cite{Lavelle2009, Li2025, Cuillel1987}. 
 
\begin{figure*}[ht]\centering
\includegraphics[width=0.9\textwidth]{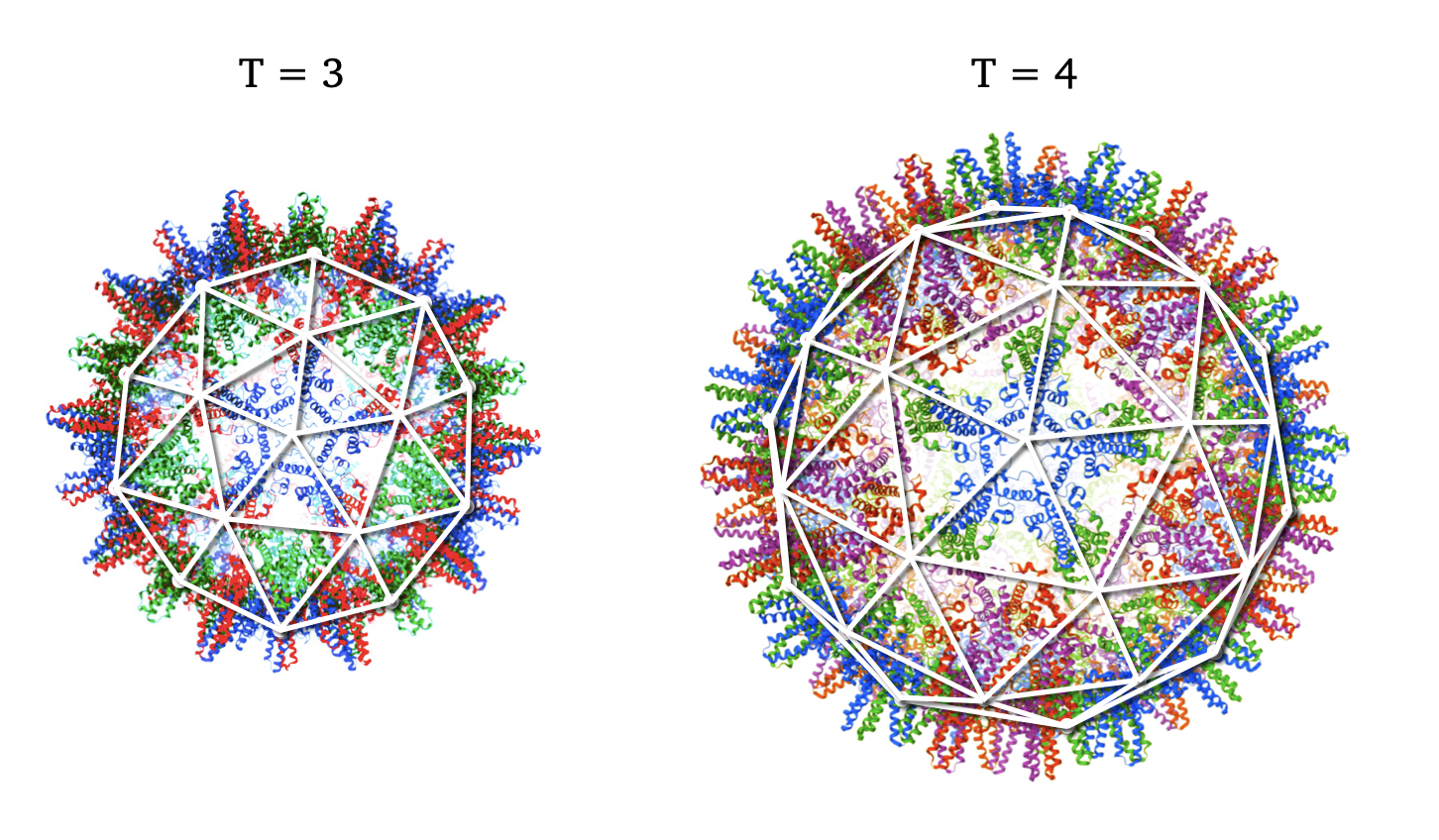}
\caption{
\textbf{Comparison of HBV capsids with $T=3$ (left) and $T=4$ (right) structures.} Both capsids are built from the same protein, with the $T=3$ structure containing 90 dimers and the $T=4$ structure 120 dimers. Icosahedral symmetry is highlighted by white auxiliary lines. Protein monomers are colored according to their symmetry-related positions within the lattice: blue marks the five monomers around each 5-fold (pentagonal) vertex, shared by both $T=3$ and $T=4$ capsids; green and red mark neighboring monomers; purple marks monomers unique to the $T=4$ geometry. Structural data: Protein Data Bank 6BVN~\cite{Schlicksup2018} ($T=3$), 8UYX~\cite{Bianchini2025} ($T=4$).} 
\label{fig:Capsid}
\end{figure*}

A well-studied example of capsid dimorphism is the hepatitis B virus (HBV), an enveloped icosahedral virus responsible for infectious liver disease~\cite{Liang2009-Hep,Block2016-HBV,Sun2018,Sun-Ionic2018}. In solution, the HBV coat proteins exist as dimers which, across a broad range of experimental and physiological conditions, can assemble into two distinct capsid geometries: $T = 3$ (comprising 90 dimers) and $T = 4$ (comprising 120 dimers)~\cite{Sun2018,Selzer2015,Asor2019,Moerman2016,dryden2006,Sheena2022}. Figure \ref{fig:Capsid} illustrates how the protein dimers are organized into the two morphologies.

Interestingly, particles formed by a dimeric truncation of the HBV capsid protein known as Cp149\textsubscript{2} have been shown to spontaneously self-assemble into both $T = 3$ and $T = 4$ structures at room temperature in the absence of genomic material, and under near-neutral pH and moderate to high salt concentrations. In the \textit{in vivo} environment, $\sim$95\% of the particles adopt $T = 4$ symmetry, while the remaining 5\% exhibit $T = 3$ symmetry. However, under \textit{in vitro} conditions, the ratio of $T = 4$ to $T = 3$ particles varies and is sensitive to thermodynamic parameters such as the salt~\cite{Moerman2016, Sun-Ionic2018, Sun2018}, pH~\cite{Moerman2016, Sun-Ionic2018, Sun2018, timmer2022}, and protein concentrations~\cite{Harms2015,Asor2019,Asor2020}. 

Several theoretical and experimental studies have shed light on factors contributing to dimorphism in HBV. Harms \textit{et al.}~\cite{Harms2015} used \textcolor{black}{resistive-pulse sensing (RPS)} to monitor viral assembly in real time, enabling detection of capsids and intermediates at concentrations well below the limits of most other techniques. From these measurements, they determined the relative abundance of $T=3$ and $T=4$ particles. Moerman \textit{et al.}~\cite{Moerman2016} developed a theoretical framework that introduced a curvature-dependent free energy term into the interaction between coat proteins, which favored the formation of $T=4$ capsids. They further analyzed the $T=4$/$T=3$ ratio as a function of protein concentration and found excellent agreement with the experimental measurements of Harms et al.~\cite{Harms2015} in the range of $0.5\mu\mathrm{M}$ to $10\mu\mathrm{M}$.

Using a coarse-grained model, Mohajerani \textit{et al.}~\cite{Mohajerani2022} simulated HBV assembly and showed that changes in the binding free energy of dimers strongly influence capsid morphology. They observe that the presence of salt in assembly environment would shift the equilibrium towards dimer states that are favoring pathways leading to T=3 capsids. These findings emphasize that coupling between dimer conformational transitions and inter-dimer interaction strengths plays a key role in guiding assembly pathways and determining the final capsid structure~\cite{Moerman2016,Harms2015}.

Using solution small-angle X-ray scattering(SAXS), Asor \textit{et al.}~\cite{Asor2019} showed that HBV capsid assembly follows a narrow pathway involving only a few hundred intermediates, enabled by an energetic bias favoring stable species. Productive assembly occurs within a narrow range of binding free energies; without energetic selection, assembly would be blocked by entropic barriers. They found that at low salt concentrations, $T = 4$ capsids dominate, while higher salt concentrations increase the proportion of $T = 3$ particles. Small changes in the binding free energy
ratio significantly shift the $T = 3$ to $T = 4$ distribution, highlighting the sensitivity of capsid polymorphism to ionic conditions. Under strong binding free energies, high temperature, and high protein concentrations, kinetically trapped $T = 3$-like nanoparticles formed, further illustrating the delicate balance
required for successful assembly.

To further probe the energy sensitivity of Cp149\textsubscript{2} assembly, Asor \textit{et al.}~\cite{Asor2020} examined the process under varying salt concentrations. Using SAXS, they tracked the time evolution of the dominant assembled species and found that the assembly pathway is largely determined by intermediates formed within the first second, even though $T=4$ capsids do not appear until about 10 seconds later. Higher salt concentrations promoted more kinetically trapped intermediates, increased the fraction of $T=3$-like particles, and reduced the relative $T=4$ population. Together, these studies show that although thermodynamics sets the driving force for Cp149\textsubscript{2} assembly, it does not determine how the process unfolds—particularly how the earliest intermediates steer the outcome toward different $T=3$/$T=4$ ratios. 

While thermodynamic models can predict the ratio of $T=3$ to $T=4$ capsids under equilibrium conditions, they assume continuous subunit exchange between assembled capsids and free dimers—an assumption not consistently supported by experiments. Uetrecht \textit{et al.}~\cite{Uetrecht2010} found that HBV capsids exhibit extremely slow subunit exchange, detectable only for $T=3$ particles at $4^\circ$C and not at physiological temperatures, even over observation periods exceeding 100 days. No exchange of $T=4$ capsid proteins was observed in their experiments.  Nevertheless, Asor \textit{et al.} showed that the initial assembly process is rapid: $T=3$ capsids form almost instantaneously~\cite{Asor2019,Asor2020}, whereas $T=4$ capsids first appear only after about 10 seconds~\cite{Asor2020}, even though $T=4$ is the predominant final assembly product. Sun \textit{et al.}~\cite{Sun2018} further showed that slowly adding salt after the onset of assembly does not alter the final distribution of capsids. Together, these experiments suggest that kinetic effects cannot be neglected in understanding capsid assembly.

These experiments highlight the role of salt in modifying the population distribution between $T=3$ and $T=4$ capsids. Although many studies have varied salt concentration to investigate capsid assembly~\cite{Sun-Ionic2018, Sun2018, Asor2020, Harms2015, Salunke1989, Mohajerani2022, Zlotnick1999-mq}, the quantitative relationship between ionic strength and subunit binding free energy remains underexplored. In particular, experiments~\cite{Sun-Ionic2018, Asor2020} demonstrated that \textcolor{black}{different} monovalent salts exert comparable influences and that the magnitude of this influence is a function of the ionic strength. These findings suggest that the observed effects are primarily governed by electrostatic screening, rather than the specific chemical identity of the ions. Of course, multivalent ions, such as calcium~\cite{SV40calcium2001,Zhou2009} and zinc ions~\cite{Khayenko2025}, may be involved in specific interactions that modulate the interaction between coat proteins, an effect that we do not consider in this paper where we focus on the electrostatic effect.

In this study, we test the scope and validity of both kinetic and thermodynamic models against experimental data. Building on the thermodynamic analysis of Moerman \textit{et al}.~\cite{Moerman2016} and explicitly including the effect of salt concentration, we examine the ratio of $T=3$ to $T=4$ capsids as a function of ionic conditions. We find that the thermodynamic model performs well within a range of moderate protein and salt concentrations, consistent with the experimental observations of Asor et al.~\cite{Asor2020}. At low protein concentrations, however, assembly is better described by a classical nucleation theory (CNT)–based kinetic model, which captures the time-dependent evolution of capsid formation \cite{Harms2015}. Notably, the kinetic model also reproduces the time evolution of assembly under moderate conditions \cite{Asor2020}, providing a more comprehensive picture of the process. To extend existing theoretical descriptions and systematically incorporate the influence of salt, we introduce an empirical formulation that quantifies changes in dimer binding free energy as a function of ionic strength. Incorporating this correction into both thermodynamic and kinetic models improves agreement with experiments across a broad range of conditions. By providing a unified treatment of salt effects, this approach connects thermodynamic and kinetic descriptions of HBV assembly and establishes a consistent basis for relating environmental conditions to viral polymorphism.

The relevance of our study extends beyond HBV. Polymorphism in viruses and other protein shells also arises in a diverse collection of biological systems, such as encapsulin, a protein-based organelle used by certain bacteria for the storage and transport of iron~\cite{Lie2023}. Eren \textit{et al.}~\cite{Eren2022} demonstrated that recombinant EncA encapsulin forms coexisting $T = 1$ and $T = 3$ cages, with their relative abundance and binding energies suggesting that dimorphism is driven by differences in subunit conformational free energy. \textcolor{black}{Similarly, ssRNA bacteriophages such as MS2 can assemble into coexisting T=3, T=4, and mixed morphologies depending on assembly conditions~\cite{Thongchol2023, deMartinGarrido2020}. In addition, a recent study on the packaging signal demonstrated that dispersed stem-loops within the gRNA cooperate to modulate nucleation and packaging selectivity, ultimately affecting capsid morphology~\cite{Rastandeh2025}.  Together, these examples highlight the broader relevance of our combined thermodynamic and kinetic model for explaining viral polymorphism across diverse self-assembling biological systems.} 

Understanding the factors that contribute to viral polymorphism may offer valuable insights for the design and control of synthetic nanocages, the development of protein-based nanomaterials, and the broader understanding of self-assembly principles in biological systems.

The remainder of this paper is organized as follows. In Section~\ref{section:models}, we present the theoretical frameworks used to construct our thermodynamic model, the kinetic model, and a proposed term to link the dimer binding free energy to salt concentration. In Section~\ref{section:Result}, we compare the performance of the thermodynamic and kinetic models with experimental data. We show that the thermodynamic model accurately predicts the relative abundance of assembled capsid structures, while the kinetic model captures the dynamic assembly of $T=3$ and $T=4$ capsids, particularly at low protein concentrations.  We discuss how both models describe HBV capsid assembly under varying salt conditions. We also analyze a key phenomenon: that salt can strongly modulate the $T=3$/$T=4$ capsid ratio when introduced before assembly, not during it. This behavior underscores the importance of kinetic effects in the capsid assembly process. 

%
%

\section{Thermodynamic and Kinetic Models} \label{section:models}
In this section, we first present the thermodynamic framework used to understand the ratio of the two HBV species, and then apply kinetic theory to analyze the pathways and rates governing their formation.

\subsection{Thermodynamic Principles}
To study how thermodynamics affects the relative abundance of $T=3$ and $T=4$ capsids, we consider a dilute solution of subunits. Given that HBV capsid proteins first associate into stable dimers in solution, which function as the primary structural units observed experimentally, we model the dimer as the fundamental building block throughout the assembly process. 

Under equilibrium conditions, subunits self-assemble into $T = 3$ or $T = 4$ capsids, containing $q_3 = 90$ and $q_4 = 120$ subunits, respectively. While intermediate or kinetically trapped capsids have been observed under certain experimental conditions, we treat these species as being in equilibrium with free dimers and include them in the free dimer concentration. This simplification, which reduces the system to three primary species in solution, is in particular supported by experimental evidence that intermediate structures during virus assembly are typically short-lived~\cite{Harms2015,Asor2020}. Therefore, we model the system as a solution containing free dimers and complete $T=3$ and $T=4$ capsids. Assuming a constant volume, the law of mass conservation would dictate that $x_\mathrm{tot} = x_\mathrm{s} + q_3 x_3 + q_4 x_4$, where $x_\mathrm{s}$ is the equilibrium concentration of free dimer subunits, and $x_3$ and $x_4$ are the concentrations of $T = 3$ and $T = 4$ capsids, respectively.

The Helmholtz free energy of the system can then be written as 
\begin{equation}\label{eqn:Helmholtz} 
\frac{\mathcal{F}}{k_\mathrm{B}T} = \sum_{i\in\{s,3,4\}} \left[ x_i \ln x_i - x_i + q_i x_i g_i \right], 
\end{equation} 
where the sum represents the contributions of free subunits $(s)$ and $T = 3$ and $T = 4$ capsids. The term $x_i \ln x_i - x_i$ represents the translational entropy of species $i$, while $q_i x_i g_i$ corresponds to the (dimensionless) binding free energy of subunits within fully formed capsids. Here, $g_i$ denotes the binding free energy difference between a fully bound subunit and a free subunit, with the reference chosen such that $g_\mathrm{s}=0$. By definition, $q_\mathrm{s}=1$.  The binding free energy incorporates contributions from Coulomb and hydrophobic interactions, hydrogen bonding, elastic energy and other effects such as the preferred curvature~\cite{Kegel2004}. Based on this formalism, we consider both $T=3$ and $T=4$ as thermodynamically stable products. 

Minimizing the free energy given in Eq.~\ref{eqn:Helmholtz} under the constraint of mass conservation that fixes $x_\mathrm{tot}$, leads to the following law of mass action relations 
\begin{equation} \label{eq:MassAction3}
x_3 = \left(\frac{x_\mathrm{s}}{c^{*}_{3}} \right)^{q_3}
\end{equation} 
and
\begin{equation} \label{eq:MassAction4}
x_4 = \left( \frac{x_\mathrm{s}}{c^{*}_{4}} \right)^{q_4}. 
\end{equation}
Here, $c_i^* = e^{g_i}$ represents the critical concentration for species $i$\cite{}. Applying the mass conservation constraint again yields an additional equation that specifies $x_s$ for a given total dimer concentration $x_\mathrm{tot}$, \begin{equation}\label{eqn:populationbyconc} 
x_\mathrm{tot} = x_\mathrm{s} + q_3 \left( \frac{x_\mathrm{s}}{c^{*}_3} \right)^{q_3} + q_4 \left( \frac{x_\mathrm{s}}{c^{*}_4} \right)^{q_4},
\end{equation} 
which describes a system in which each dimer can exist in one of three states: (i) as a free subunit, (ii) incorporated into a $T=3$ capsid, or (iii) incorporated into a $T=4$ capsid.  It has been shown in previous studies that obtaining exact analytical solutions to these equations is nontrivial~\cite{Zandi2009,Clark2023}. 

In the model of Eq.~\ref{eqn:populationbyconc}, capsid assembly is assumed to be fully reversible. Experimental studies, however, show that once capsids have formed, they do not instantaneously disassemble into free dimers even when conditions are reversed to favor disassembly~\cite{Harms2015,Uetrecht2010,Sun-Ionic2018,Singh2003}. This irreversibility gives rise to hysteresis, where the assembly and disassembly pathways depend on the system’s prior history~\cite{Starr2022,Singh2003}.In such cases, the system does not retrace the same trajectory when external parameters (e.g., concentration or interaction strength) are reversed, but instead follows a distinct route that reflects the presence of kinetic barriers and metastable states. As a result, capsids that are thermodynamically unfavorable can nonetheless remain assembled, stabilized temporarily by kinetic trapping. These kinetic bottlenecks and alternate pathways play a crucial role in determining which morphologies ultimately appear. In the following section, we investigate how such kinetic effects govern assembly outcomes and contribute to the coexistence of multiple capsid species \cite{Singh2003,timmer2022}.

\subsection{Dynamic Modeling of Viral Capsid Assembly}

We describe the kinetics of capsid growth using CNT~\cite{Zandi2006,zandi2020virus,Hagan2011}, in which the system begins as a solution of free subunits, and assembly proceeds by first overcoming a free-energy barrier to form a critical nucleus. Once this nucleus is established, additional subunits are added rapidly, and the process can be understood by analyzing the free-energy landscape associated with intermediate structures of varying sizes. The Gibbs free energy $\Delta G(n)$ for a partially assembled capsid containing $n$ subunits reflects a competition between favorable bulk interactions, which promote subunit addition, and unfavorable edge contributions, which resist growth due to incomplete boundaries. The free energy is given by
\begin{equation}
\label{eqn:energybarrier}
\Delta G(n) = n \Delta \mu + \frac{4 \pi R\, \sigma}{q} \sqrt{n(q - n)},
\end{equation}
where $\Delta \mu \equiv \mu_a - \mu_f$ is the chemical potential difference between subunits in the assembled state ($\mu_a$) and the free state ($\mu_f$). When $\Delta \mu < 0$, the first term favors growth by lowering the system’s free energy as subunits are incorporated. The second term captures the energetic cost of maintaining an exposed rim in a partially formed shell, where $q$ is the number of subunits in a complete capsid and $R$ is its effective radius. The line tension $\sigma$, which governs this edge penalty, is given by $\sigma = -\frac{s g \sqrt{q}}{4R}$, where $s$ is a geometric factor (ranging between $0$ and $1$), $g$ is the dimer binding free energy within the capsid, and $R$ is the effective radius of the shell~\cite{Zandi2006}.

According to Becker–Döring kinetics, which describes nucleation through the sequential attachment and detachment of single subunits, the steady-state nucleation rate per unit volume, $J$, is given by \cite{Becker1935},
\begin{equation}
\label{eqn:flux}
J = x_s \nu Z \exp\left[-\Delta G(n^*)\right],
\end{equation}
where $x_s$ is the concentration of free subunits, $\nu$ corresponds to the attempt frequency of dimers attaching to the critical nucleus, and $n^*$ is the critical nucleus size at which the free energy $\Delta G(n)$ reaches its maximum. The Zeldovich factor, $Z=\sqrt{\frac{-\Delta G''(n^*)}{2\pi}}$, quantifies the curvature of the free energy barrier near $n^*$. The height of the energy barrier, $\Delta G(n^*)$, dominates the exponential term and thus strongly influences the assembly rate.
Within an adiabatic approximation, the time evolution of capsid populations can then be written by the following system of differential equations:
\begin{equation}
\label{eq:rate}
\begin{cases}
\frac{d x_{\mathrm{s}}}{dt} = -J_3 q_3 - J_4 q_4 \\
\frac{d x_3}{dt} = J_3 \\
\frac{d x_4}{dt} = J_4.
\end{cases}
\end{equation}
\noindent Here, $x_{\mathrm{s}}$ denotes the concentration of free dimers, while $J_3$ and $J_4$ are the assembly rates of $T=3$ and $T=4$ capsids, respectively, as defined by Eq.~\ref{eqn:flux}. Each $T=3$ capsid incorporates $q_3 = 90$ dimers, and each $T=4$ capsid uses $q_4 = 120$ dimers. As the capsid assembly progresses, $x_{\mathrm{s}}$ decreases, while the number of completed capsids $x_3$ and $x_4$ increases. We numerically solve the differential equations given in Eq.~\ref{eq:rate} using a forward Euler method with adaptive time steps, following the approach of Timmermans \textit{et al.}~\cite{timmer2022}. We set the initial conditions such that $x_{\mathrm{s}}$ equals the initial free dimer concentration, with $x_3 = x_4 = 0$. Because $J_3$ and $J_4$ can differ significantly due to their exponential sensitivity to the (free) energy barrier, we use an adaptive time step, $dt=k \min(x_\mathrm{tot}/J_3,x_\mathrm{tot}/J_4)$ where $x_{\text{tot}}\equiv x_{\mathrm{s}}+q_{\mathrm{3}} x_{\mathrm{3}}+q_{\mathrm{4}} x_{\mathrm{4}}$ is the total dimer concentration and $k$ is a tuning parameter that sets the time resolution. Unless otherwise noted, we use $k = 10^{-4}$ in all kinetic simulations.

\subsection{Quantifying the Effect of Electrostatic Screening}
Ionic conditions play a crucial role in the assembly of capsids and other supramolecular structures composed of charged molecular subunits~\cite{wang2022, Liu2012, Kegel2004}.  Salt alters subunit binding free energy by screening and thereby weakening electrostatic interactions between charged residues~\cite{Sun2018}. This effect can be captured within the Debye–H\"{u}ckel (DH) approximation, which provides a simple yet effective way to estimate the electrostatic contribution to the free energy of assembly. We incorporate this contribution into both our thermodynamic and kinetic models through a DH-based free-energy formulation, and compare the resulting predictions with experimental observations. Within this approximation, the potential of mean force between two unit charges located at positions $\vec{r}$ and $\vec{r},'$ is given by~\cite{Verway1948},
\begin{equation}
V_C = k_B T \frac{\lambda_B}{|\vec{r} - \vec{r}'|} \exp\left( -\kappa |\vec{r} - \vec{r}'| \right),
\end{equation}
where $\lambda_B = e^2/(4\pi \varepsilon_0 \varepsilon_r k_B T)$ is the Bjerrum length, $\varepsilon_0$ is the vacuum permittivity, and $\varepsilon_r$ is the relative dielectric constant of the solvent. The Debye screening length is given by $\kappa^{-1} = 1/\sqrt{8\pi \lambda_B I}$, where $I$ is the ionic strength of the solution. The ionic strength for a general electrolyte is defined as:
\begin{equation}\label{eq:ionicstrength}
I = \frac{1}{2} \sum_i c_i \zeta_i^2,
\end{equation}
where $c_i$ is the concentration of the ion species $i$ and $\zeta_i$ is its valency. 
Kegel and van der Schoot~\cite{Kegel2004} calculated the total electrostatic interaction energy between subunits of a capsid, assuming that the charge is uniformly distributed over a spherical shell of radius $R$ and thickness $d \ll R$, see also \v{S}iber \textit{et al.}~\cite{siber2012energies}   The resulting expression for the Coulomb free energy is:
\begin{equation} \label{eq:freepaul}
G_C \approx \frac{1}{4} k_B T q^2 z^2 R^{-2} \lambda_B \kappa^{-1},
\end{equation}
where $q$ is the number of subunits in the capsid and $z$ is the charge per subunit. This expression shows that the electrostatic contribution to the free energy scales linearly with the Debye length, which itself varies as $\kappa^{-1} \propto 1/\sqrt{I} \propto 1/\sqrt{[\mathrm{salt}]}$ with $[\mathrm{salt}]$ representing the molar concentration of a salt dissolved in the solution. 
Using Eq.~\ref{eq:freepaul}, we can write the free energy per subunit as follows,
\begin{equation} \label{eq:FreeEnergyVersusSalt}
g_{i,\mathrm{salt}} = g_{i,0} + \frac{b_i}{\sqrt{[\mathrm{salt}]}},
\end{equation}
where $g_{i,0}$ is the intrinsic dimer free energy in the absence of salt, and $b_i$ is a parameter that quantifies the electrostatic contribution to subunit–subunit interactions in $T=i$ capsids. The value $b_i$ depends on the temperature $T$, dielectric constant $\varepsilon_r$, capsid radius $R$, and subunit number $q$.
Although 1:1 electrolytes (e.g., NaCl) are commonly used to promote assembly, 2:1 electrolytes like calcium chloride (CaCl\textsubscript{2}) are sometimes employed for their stronger screening effect. Ignoring the possibility of specific calcium binding sites, we treat calcium as a divalent salt, which contributes a factor of $\sqrt{3}$ relative to a 1:1 salt of equal molarity. The Debye–Hückel approximation becomes less accurate in the presence of multivalent ions, where nonlinear effects become significant. In such cases, more sophisticated models, such as the full Poisson–Boltzmann theory, are preferred~\cite{Trefalt2013,Montes2014}. Nonetheless, for the purposes of our analysis of HBV data, the Debye–Hückel model provides a sufficiently accurate estimate of the salt-dependent free energy, as demonstrated in the Results and Discussion sections.

%
%

\section{Results and Discussion}\label{section:Result}

\begin{figure}
\includegraphics[width=0.85\linewidth]{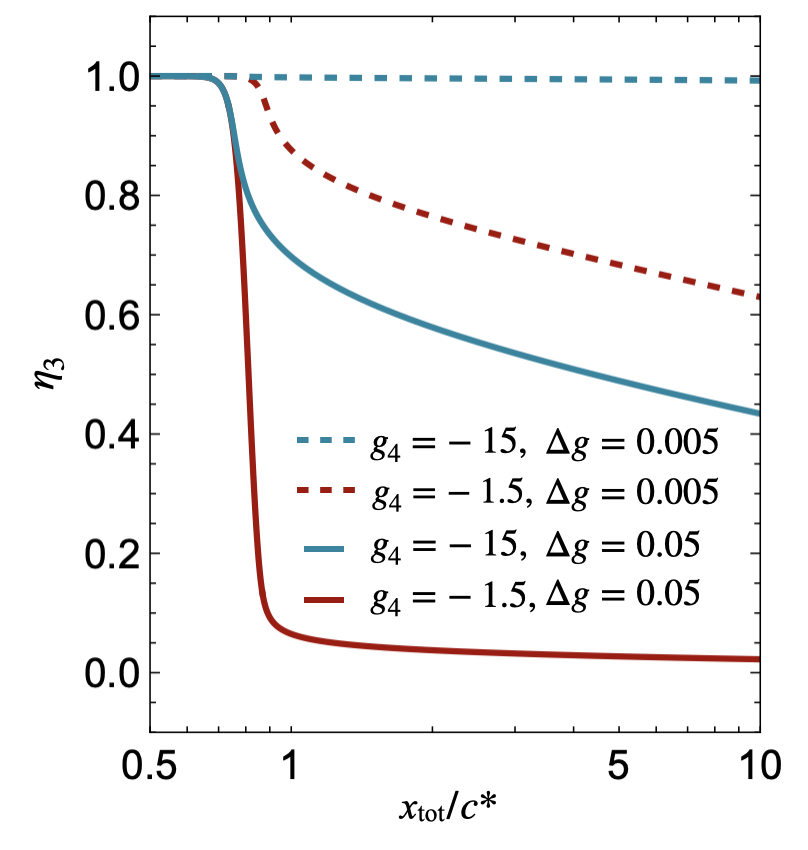}
\centering
\caption{ \textbf{
Binding free energy controls the relative abundance of $T=3$ and $T=4$ capsids.}  
The population balance between $T=3$ and $T=4$ morphologies is governed by the binding free energy $g_4$ and the energy difference $\Delta g \equiv g_3 - g_4$. At subunit concentrations near $0.5c^*$, thermodynamics predicts that nearly all assembled capsids are $T=3$. As the total subunit concentration increases, the fraction of $T=4$ capsids rises. A larger $\Delta g$ decreases $T=3$ relative abundance. \textcolor{black}{A more negative value of $g_4$ at a fixed $\Delta g$ biases the system toward $T=3$ assembly because the relative energy difference becomes smaller.}}
\label{fig:ThermoPrediction}
\end{figure}

\subsection{Binding Free Energy Differences Determine the Equilibrium Ratio of \textit{T=3} to \textit{T=4} Capsids}

To elucidate the physical principles governing the relative concentration of $T=3$ and $T=4$ capsids, we examine how differences in dimer binding free energies define the capsid distribution. Our analysis shows that the free energy difference between dimers incorporated into $T=3$ and $T=4$ capsids plays a central role in determining the final structural ratio, consistent with the experiments of Asor \textit{et al.}~\cite{Asor2019}. To quantify the fraction of dimers incorporated into capsids of different $T$ numbers, we apply Eqs.~\ref{eq:MassAction3}, \ref{eq:MassAction4}, and~\ref{eqn:populationbyconc} to determine the concentrations of $T=3$ and $T=4$ capsids, and define $\eta_i={x_i q_i}/({x_3 q_3 + x_4 q_4})$. Here, $\eta_i$ represents the fraction of all dimers in assembled capsids that are incorporated into capsids of structure $T=i$ for the cases $i=3$ and $i=4$. 

Figure~\ref{fig:ThermoPrediction} shows $\eta_3$ as a function of the total dimer concentration ($x_{\text{tot}}$, see Eq.~\ref{eqn:populationbyconc}). Four different scenarios are illustrated, highlighting the effect of the binding free energy difference $\Delta g \equiv g_3 - g_4$ and the magnitude of the free energy $g_3$ or $g_4$ in modulating the capsid distribution.
For example, when $\Delta g = 0.005\,k_\mathrm{B}T$, changing $g_4$ from $-15\,k_\mathrm{B}T$ to $-1.5\,k_\mathrm{B}T$ causes only a modest change in $\eta_3$ (from $1$ to $0.6$), with the majority of dimers still assembling into $T=3$ structures. In contrast, increasing the energy difference to $\Delta g = 0.05\,k_\mathrm{B}T$ significantly alters the structure ratio: for $g_4 = -15\,k_\mathrm{B}T$, $\eta_3$ drops to $0.5$, and for $g_4 = -1.5\,k_\mathrm{B}T$, it falls further to $0.05$. These results, obtained at total concentrations up to $10\,c^*$, indicate that the distribution of assembled capsids is governed by both the relative binding energies between dimer types and their absolute binding strengths. In the dimorphic assembly system, the critical concentration is $c^*=\min(e^{g_3},e^{g_4})$, which reflects the fact that the morphology with the stronger binding free energy (i.e. lower threshold concentration) determines the onset of assembly. 

\begin{figure*}
\centering
\includegraphics[width=1.0\linewidth]{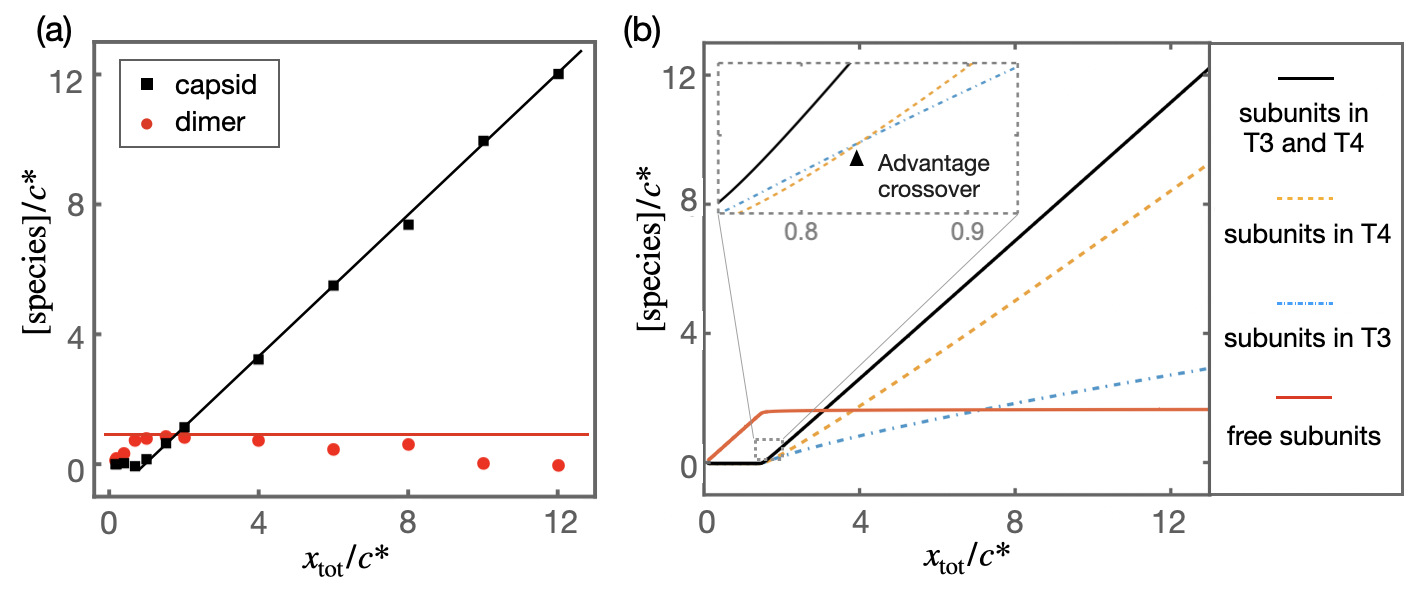}
\caption{ 
\textbf{Thermodynamic prediction of capsid assembly.} 
\textbf{(a)}  Experimental data from Harms et al.~\cite{Harms2015} showing concentrations of free dimers (red) and assembled capsids (black) versus total dimer concentration. Solid lines from the thermodynamic model are consistent with the results of Moerman et al.~\cite{Moerman2016}.
\textbf{(b)} Model predictions for $T=3$, $T=4$, and free dimers as a function of total protein concentration ($x_{\mathrm{tot}}/c^*$) reveal a critical threshold above which assembly increases sharply. Parameters fitted to experimental data are $g_4=-18.50 \,k_\mathrm{B}T$ and $\Delta g=g_{3}-g_{4}=0.07\,k_\mathrm{B}T$~\cite{Moerman2016}. The inset shows the regime where $T=4$ capsids overtake $T=3$, despite low overall equilibrium yields.
\label{fig:assemblybyconc}}
\end{figure*}

\begin{figure}[ht]
\includegraphics[width=0.45\textwidth]{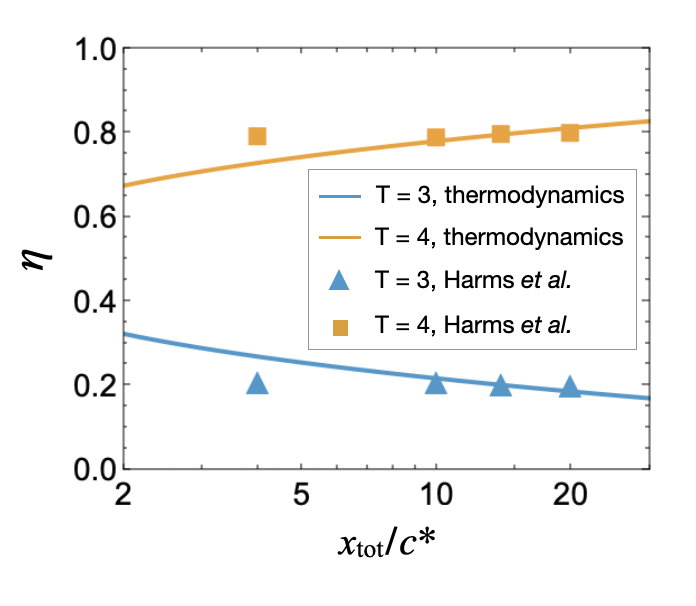}
\caption{
\textbf{Comparison of theory and experiment at moderate concentrations.} Relative populations of $T=3$ (blue triangles) and $T=4$ (orange squares) from Harms \textit{et al.}~\cite{Harms2015} vs.~the total dimer concentrations ($x_{\mathrm{tot}}/c^*$). Solid lines show thermodynamic predictions with parameters $g_4=-18.50k_\mathrm{B}T$ and $\Delta g=0.07k_\mathrm{B}T$.}
\label{fig:Zlotinick}
\end{figure}
The thermodynamic model can also successfully predict the assembly behavior of HBV capsids, most notably the existence of a critical concentration. Figure~\ref{fig:assemblybyconc}a compares our thermodynamic model predictions with the experimental data~\cite{Harms2015}, showing capsid concentrations as a function of total dimer concentration. The square symbols represent the total capsid concentrations, while the red dots indicate the free dimer concentrations measured experimentally. The solid lines show the corresponding results from our thermodynamic model (see Eqs.~\ref{eq:MassAction3}-\ref{eqn:populationbyconc}). Figure \ref{fig:assemblybyconc}b presents the theoretical concentrations of $T=3$ and $T=4$ capsids, along with free dimers. Note that the experiments report only the concentration of assembled capsids, without distinguishing between $T=3$ and $T=4$ structures. The critical concentration, $c^*=0.5$ mM, is estimated from the concentration of free dimer remaining in the equilibrated system~\cite{Harms2015}. 

Our model predicts a crossover point at approximately $0.8\,c^*$: below this concentration, $T=3$ capsids are slightly more prevalent, whereas at higher concentrations $T=4$ becomes dominant (see the inset of Fig.~\ref{fig:assemblybyconc}b). This crossover corresponds to the concentration at which equal fractions of dimers are incorporated into $T=3$ and $T=4$ capsids, and can be determined by equating Eq.~\ref{eq:MassAction3} and Eq.~\ref{eq:MassAction4}:
\begin{equation}
	\label{eqn:crossover}
	x_{\text{free,cx}}=\left(\frac{q_3}{q_4}\frac{\left(c_4^*\right)^{q_4}}{\left(c_3^*\right)^{q_3}}\right)^{\frac{1}{q_4-q_3}}.
\end{equation}
We note that, by definition, the number of dimers in a complete capsid satisfies $q \gg 1$, with $q_4 > q_3$. Considering that the critical concentration is $c^*_i=e^{g_i}$, the crossover point scales exponentially with $g_4+\frac{q_3}{q_4 - q_3}(g_4-g_3)$. Moerman \textit{et al.}~\cite{Moerman2016} showed that dimorphism can only occur when $g_4\leq g_3$. Therefore, the term in parentheses in Eq.~\ref{eqn:crossover} must be less than one. Thus the crossover point predicted by thermodynamics always lies below the critical concentration. 

We add a cautionary note that the thermodynamic model effectively describes capsid assembly only within a moderate range of subunit concentrations. Below $1\,c^*$, it predicts negligible assembly; above $20\,c^*$, experimental studies report aggregation phenomena not captured by the model. These limitations suggest that the model should be applied with care outside this concentration window. 

Within the concentration range $1c^*$ to $20c^*$, thermodynamic predictions are in close agreement with experimental observations. Figure~\ref{fig:Zlotinick} illustrates this agreement by plotting the yield of $T = 3$ and $T = 4$ capsids, denoted $\eta_3$ and $\eta_4$, as functions of total dimer concentration. The data points represent experimental measurements from Harms \textit{et al.}~\cite{Harms2015}, while the solid lines are obtained from the equilibrium thermodynamic model.  All parameters are provided in the figure caption.  The model captures the relative proportions of $T=3$ and $T=4$ particles across the measured range. The quantitative agreement with experiments shows that equilibrium thermodynamics reliably describes polymorphic capsid assembly at intermediate concentrations. In this regime, the ratio of $T=3$ to $T=4$ capsids can be directly obtained by dividing the equilibrium mass-action expressions in Eqs.~\ref{eq:MassAction3} and~\ref{eq:MassAction4}, 
\begin{equation}
\label{eqn:ratiothermo}
\frac{x_3}{x_4}=\exp{\left[-q_4 \left( g_3-g_4\right) \right]}.
\end{equation}
Since $q_3$ and $q_4$ are constants, Eq.~\ref{eqn:ratiothermo} shows that even small differences in subunit free energy ($\Delta g = g_3 - g_4$) can lead to large shifts in the population ratio, emphasizing the model’s sensitivity to these parameters. \textcolor{black}{We note that although our findings focus on HBV capsid assembly and \textit{in vitro} experiments performed in the absence of genomic RNA, the presence of gRNA can significantly influence both the nucleation barrier and the effective binding free energy between dimers~\cite{Patel2017,Patel2021,Li2017,Li2022}.}

\subsection{Influence of Salt Concentration on Dimer Binding Free Energy}

\begin{figure*}
\centering
\includegraphics[width=17.0cm]{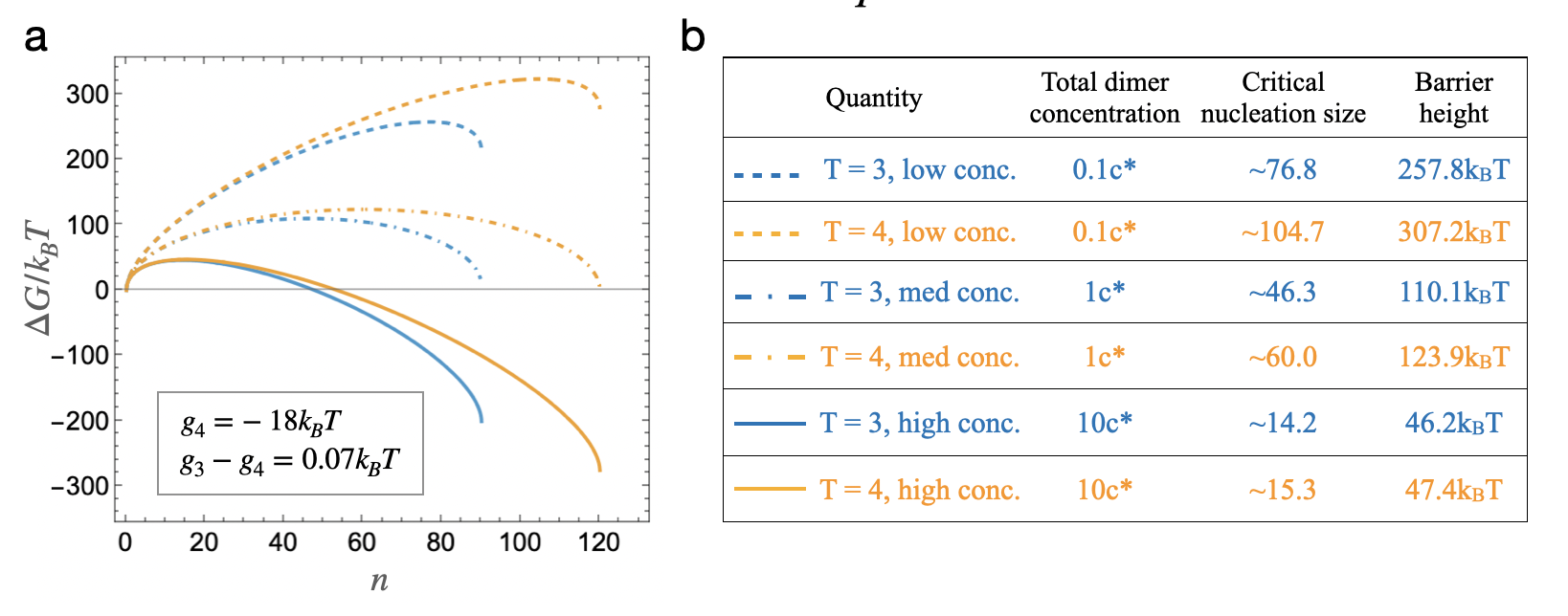}
\caption{\textbf{Free-energy profiles of capsid assembly vs $n$.} 
\textbf{(a)} Free-energy profiles for $T=3$ (blue) and $T=4$ (orange) capsids at different total dimer concentrations. At higher concentrations, both the barrier height and the critical nucleation size decrease. \textcolor{black}{Parameters were chosen as $g_4 = -18.00\,k_\mathrm{B}T$ and $g_3 - g_4 = 0.07\,k_\mathrm{B}T$, values similar to those reported in} Moerman~\cite{Moerman2016}.
\textbf{(b)} Tabulated values of the critical nucleation size and barrier height, numerically calculated at multiple concentration.}
\label{fig:fig_energybarrier}

\end{figure*}

Salt concentration plays a critical role in viral assembly by shifting the balance between competing capsid structures. In the case of HBV, variations in salt concentration have been observed to change the relative abundance of $T=3$ and $T=4$ capsids. Sun \textit{et al.}~\cite{Sun-Ionic2018} systematically examined the assembly of HBV Cp149\textsubscript{2} at varying salt concentrations and found that higher concentrations favor the formation of $T=3$ capsids. In their experiments, samples were prepared at different CaCl\textsubscript{2}	concentrations, allowed to assemble until equilibrium, and then analyzed by size-exclusion chromatography to determine the concentrations of dimers as well as $T=3$ and $T=4$ capsids. From these equilibrium concentrations, the corresponding equilibrium constants were extracted, which in turn enabled calculation of the dimer binding free energy according to
$g_\text{contact} = -N_\mathrm{A} k_\mathrm{B}T\ln(K_\text{contact})$,
where $N_\mathrm{A}$ is Avogadro’s number and $K_\text{contact}$ is obtained from the experimentally determined equilibrium constants (see supplementary material B for derivation details).
Fitting the experimental data of Sun \textit{et al.}~\cite{Sun2018} with the relationship introduced in Eq.~\ref{eq:FreeEnergyVersusSalt}, we obtained the following parameters:
$g_{3,0} = -26.000\,k_\mathrm{B}T$,
$g_{4,0} = -26.025\,k_\mathrm{B}T$,
$b_3 = 18.755\,k_\mathrm{B}T\,\mathrm{mM}^{1/2}$\,
and $b_4 = -18.327\,k_\mathrm{B}T\,\mathrm{mM}^{1/2}$ (See Fig~S1 in the supplementary material).
At $20,\mathrm{mM}$ CaCl\textsubscript{2}, our fit yields a binding free energy difference between $T=3$ and $T=4$ of approximately $0.2\,k_\mathrm{B}T$, comparable in magnitude to the curvature energy differences estimated by Moerman \textit{et al.}~\cite{Moerman2016} and Asor \textit{et al.}~\cite{Asor2020}.
The fitted parameters reveal that the electrostatic sensitivity coefficient, $b_i$, is marginally larger for $T = 3$ dimers, indicating that their binding free energy exhibits a stronger dependence on ionic strength. This suggests enhanced electrostatic screening for $T = 3$ relative to $T = 4$, potentially reflecting differences in local charge distribution or subunit interface geometry.

Our fitted values of dimer binding free energy are consistent with experimentally reported estimates, which range from 20 to 28  $k_\mathrm{B}T$~\cite{Asor2019, Harms2015}. Although the differences in $g_{i,0}$ and $b_i$ between $T = 3$ and $T = 4$ structures appear modest, their implications are non-negligible. 

In the following sections, we quantitatively demonstrate that both the thermodynamic equilibrium and kinetic assembly models are highly sensitive to the values of $g_{i,0}$ and $b_i$. Even small deviations in intrinsic binding free energy ($g_{i,0}$) or in the electrostatic sensitivity coefficient ($b_i$) can lead to significant shifts in the relative stability and prevalence of the resulting polymorphic capsid products.

\subsection{Kinetics of Capsid Polymorphism at Low Subunit Concentrations}

\begin{figure*}[tb]

\centering
\includegraphics[width=0.95\linewidth]{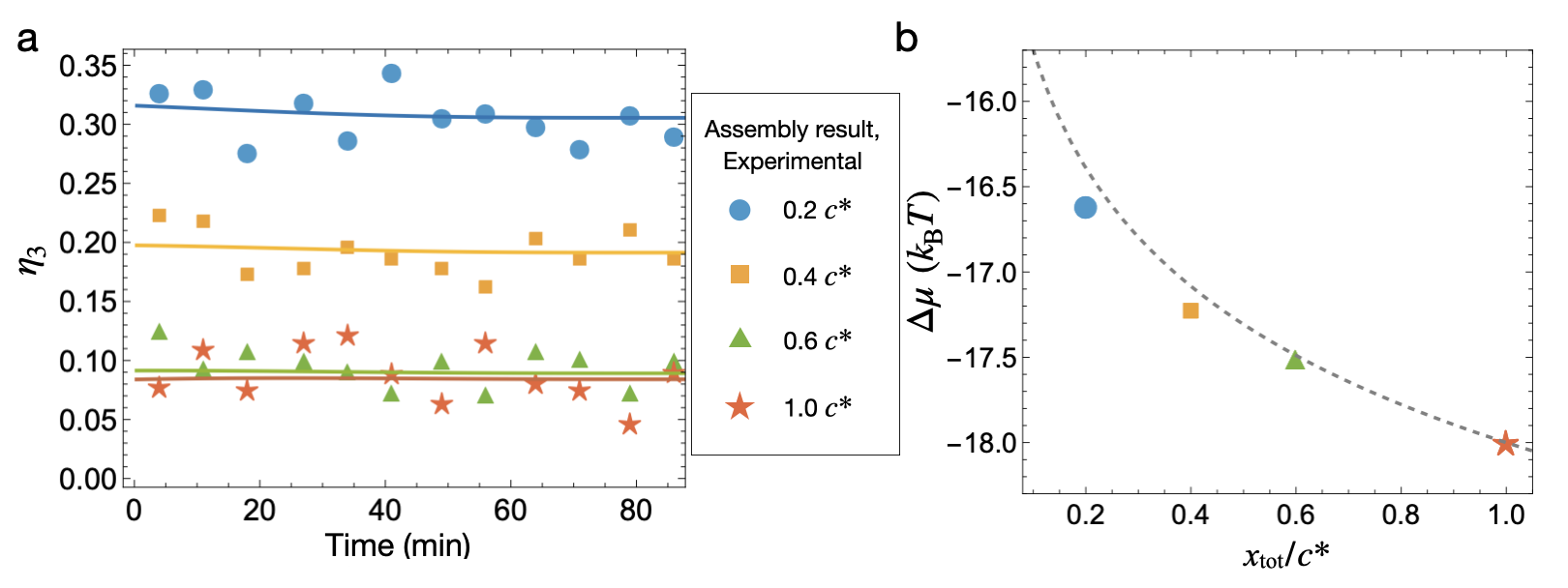}
\caption{ 
\textbf{Dynamic capsid assembly behavior at and below critical concentration.} 
\textbf{(a)} Sub-critical capsid assembly behavior from Harms \textit{et al.}~\cite{Harms2015}. Blue circles, orange squares, green triangles, and red stars correspond to total dimer concentrations of 0.1, 0.2, 0.3, and 0.5 \textmu M, respectively. Solid lines show kinetic model predictions of $\eta_3$ as a function of time, with matching colors.
\textbf{(b)} Symbols show the $\Delta \mu$ values used as fitting parameters in part (a), plotted against total protein concentration. The dashed line, $-18k_{\mathrm{B}}T - k_{\mathrm{B}}T\ln(x_\text{tot}/c^*)$, indicates that under sub-critical conditions $\Delta \mu$ decreases logarithmically with concentration. Notably, the offset is the $T=4$ dimer binding free energy $g_4$.
\label{fig:fig_low_cont_kinetics}}
\end{figure*}

In the sections above, we established a thermodynamic model for predicting the equilibrium distribution of dimers among polymorphic capsids. In many experimental situations, however, kinetics may play a more important role than thermodynamics, since the global free-energy minimum is not necessarily the state most accessible through kinetic pathways. Notably, the kinetic effects successfully account for the experimentally observed hysteresis in capsid assembly~\cite{Singh2003}. In particular, dissociation of fully formed capsids begins as the total protein concentration is lowered below the critical threshold -- a regime in which thermodynamic models predict no capsid assembly should occur.

To explore assembly at low concentrations, we adopt a kinetic framework based on CNT to model the process of capsid assembly. Under this framework, Eq.~\ref{eqn:flux} states that the rate of assembly is governed by the height of the free energy barrier. In Fig.~\ref{fig:fig_energybarrier}a, we illustrate $\Delta G$ as a function of the number of dimers in a $T=4$ capsid for different values of total dimer concentration. The height of the energy barrier depends on dimer concentration, even when \textcolor{black}{the binding free energies $g_4=-18.00\,k_{\mathrm{B}} T$ and $\Delta g=0.07\, k_{\mathrm{B}} T $ are held constant} (see Fig.~\ref{fig:fig_energybarrier}b). This dependence arises from the chemical potential term in Eq.~\ref{eqn:energybarrier}, which is directly tied to the dimer concentration.

To study the experiments of Harms \textit{et al.}~\cite{Harms2015}, in which the fraction of subunits in capsids, $\eta$, is measured as a function of time at and below the critical dimer concentration (Fig.~\ref{fig:fig_low_cont_kinetics}a), we numerically integrate the rate equations given in Eq.~\ref{eq:rate} to retrieve the capsid assembly dynamics. Different symbols in the figure represent experiments performed at different total protein concentrations. To obtain the time evolution of capsid concentrations, we first fit the data at the critical dimer concentration, where the equilibrium condition $\Delta \mu = 0$ applies. In our model, $T=3$ and $T=4$ capsids differ only in the parameters $g$ and $q$. To fit the data at low concentrations from Harms \textit{et al.}~\cite{Harms2015}, we fixed $g_4 = -18\,k_\mathrm{B}T$, consistent with previous theoretical and experimental studies~\cite{Moerman2016,Asor2020}, and adjusted $g_3$. From the relative concentrations of $T=3$ and $T=4$ at the critical dimer concentration $c^*$ (Fig.~\ref{fig:fig_low_cont_kinetics}a), the kinetic model \textcolor{black}{predicts} $\Delta g = 1.24\,k_\mathrm{B}T$. Analysis of the experimental data below $c^*$ further indicates that $\Delta \mu$ can no longer be taken as zero but must vary, as expected, since the system is not in equilibrium under subcritical assembly conditions. 

Above the critical dimer concentration, all additional dimers assemble into capsids, leaving behind a constant free-dimer population equal to $c^*$. At subcritical concentrations, however, this condition no longer applies. In this regime, the chemical potential difference $\Delta \mu$ depends explicitly on the total protein concentration because ${x_3}/{c_{3}^{*}} \rightarrow 0$ and ${x_4}/{c_{4}^{*}} \rightarrow 0$ in Eq.~\ref{eqn:populationbyconc} as so few dimers form capsids at sub-critical concentrations. We fit the experimental data (symbols) in Fig.~\ref{fig:fig_low_cont_kinetics}a by treating $\Delta \mu$ as a fitting parameter, reflecting the assumption that most capsids are kinetically trapped below $c^*$; the solid lines show the theoretical predictions. The agreement between experiment and theory is remarkably good. Figure~\ref{fig:fig_low_cont_kinetics}b shows the corresponding plot of $\Delta \mu$ versus the ratio of total protein concentration to the critical concentration ($x_{\mathrm{tot}}/c^*$). A fit to the data shows $\Delta \mu = -18\, k_\mathrm{B}T - k_\mathrm{B}T \ln{\left(x_\mathrm{tot}/c^* \right)}$ (dashed line) showing that $\Delta \mu$ decreases logarithmically with protein concentration, consistent with the difference between the binding free energy $g_4 = -18\,k_\mathrm{B}T$ of a protein subunit in a $T=4$ capsid and the chemical potential of free proteins.

\subsection{Kinetic Modeling of Capsid Assembly at High Subunit Concentrations and Physiological Ionic Strengths}

\begin{figure}
\centering
\includegraphics[width=\linewidth]{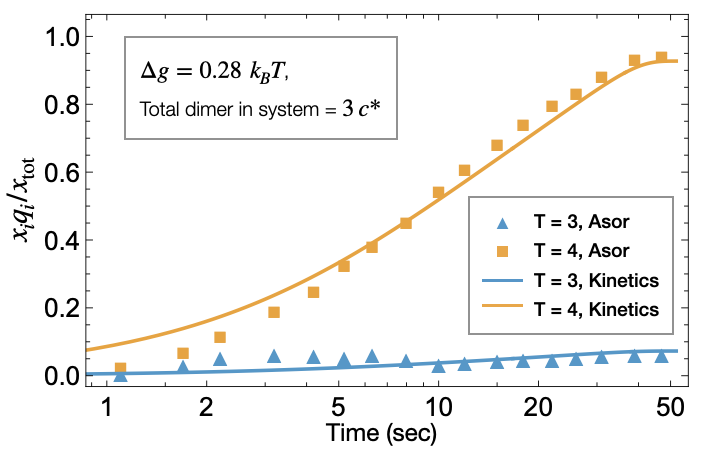}
\caption{ 
\textbf{Capsid assembly above critical concentration.}
Symbols show experimental measurements of the fraction of dimers in $T=3$ (blue triangles) and $T=4$ (orange squares) capsids. Solid curves are theoretical predictions, which closely match the data of Asor \textit{et al.}~\cite{Asor2020}.
\label{fig:highconc_kinetics}}
\end{figure}

To test the validity of our kinetic model beyond the low-concentration regime, we apply it to the experimental system of Asor \textit{et al.}~\cite{Asor2020}, who measured the normalized capsid concentration $x_i q_i / x_{\text{tot}}$ as a function of time at a total dimer concentration of $3\,c^*$. The experimental data are marked with respective symbols in Fig.~\ref{fig:highconc_kinetics}, where the solid lines represent the solutions of our kinetic model, obtained from Eq.~\ref{eq:rate} using $\Delta g = 0.28\,k_\mathrm{B}T$ as the free energy difference between $T=3$ and $T=4$ capsids. The agreement between our theoretical solutions and the experimental data is excellent, capturing both the overall time-dependent behavior and final yields of the assembly process. 

It should be noted that our model assumes an instantaneous completion of capsid formation once the critical nucleus is reached, as encoded in the flux expression (Eq.~\ref{eqn:flux}). In reality, however, free dimers first assemble into intermediate states and may experience a finite delay before completing the capsid~\cite{Asor2020, Harms2015}. This likely accounts for the slight overestimation of $T=4$ capsid formation early in the simulation and the underestimation at later stages, as seen in Fig.~\ref{fig:highconc_kinetics}.
Overall, the ability of the kinetic model to accurately reproduce the experimental time evolution—without the need for additional fitting parameters—demonstrates its applicability across a broad range of concentrations and assembly conditions. 

\subsection{Kinetic Models Better Describe Ionic-Strength Dependence in Capsid Assembly}

\begin{figure}[t]
\centering
\includegraphics[width=\linewidth]{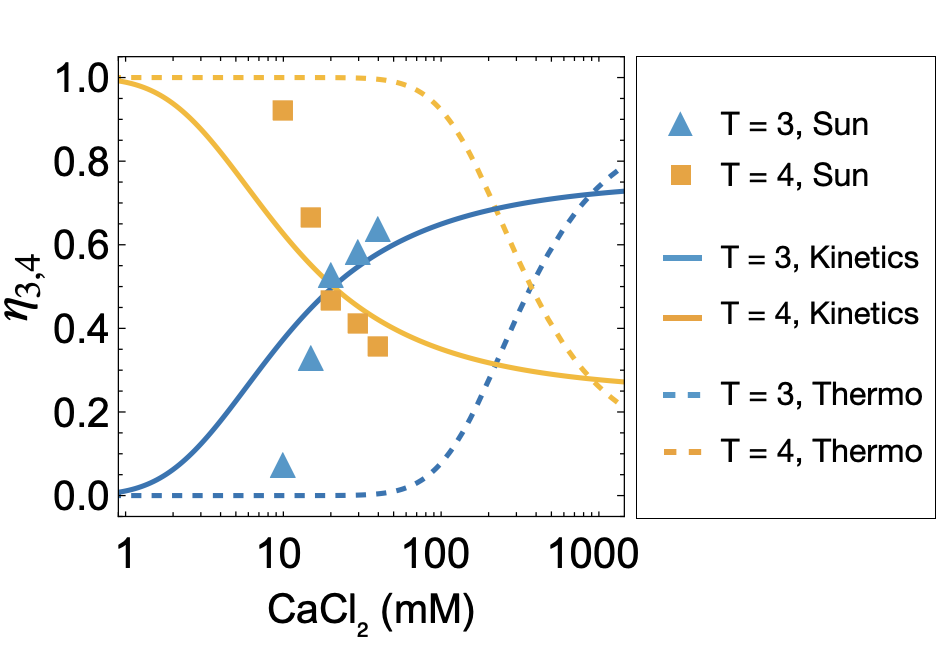}
\caption{ 
\textbf{Relationship between salt concentration and capsid structure distribution.}
Symbols show experimental data from Sun \textit{et al.}~\cite{Sun-Ionic2018}, where $T=3$ overtakes $T=4$ above $20\mathrm{mM}$ CaCl\textsubscript{2}. Solid curves are theoretical predictions using fitted free binding energies $g_3=-26.025+18.755/\sqrt{[\mathrm{CaCl_2}]}$ and $g_4=-26.000+18.327/\sqrt{[\mathrm{CaCl_2}]}$, with a total dimer concentration $10c^*$. The dashed lines show predictions from the thermodynamic model, which reproduce the trend of the experiments, but cannot capture the transition as precisely as the kinetic model. 
\label{fig:fig_salt-energy}}
\end{figure}

In the previous sections , we showed that the distribution of HBV capsid morphologies is determined by the relative free energies of competing structures. Here, we focus on experiments investigating the effect of salt on capsid assembly. Specifically, we assess how the salt-dependent binding free energy, as expressed in Eq.~\ref{eq:FreeEnergyVersusSalt}, aligns with experimental observations and whether kinetic or thermodynamic models better capture this dependence.
Experimental studies have consistently shown that increasing ionic strength biases capsid assembly toward the smaller $T=3$ morphology~\cite{Sun2018, Harms2015, Asor2020, Uetrecht2010-Natchem}. In particular, Sun \textit{et al.}~\cite{Sun-Ionic2018} reported a concentration-dependent increase in the fraction of $T=3$ capsids with rising CaCl\textsubscript{2} levels.

The experimental data of Sun \textit{et al.} presented in Fig.~\ref{fig:fig_salt-energy} illustrate the salt dependence of capsid morphologies. Symbols denote the measured molar fractions of $T=3$ ($\eta_3$) and $T=4$ ($\eta_4$) capsids as functions of CaCl\textsubscript{2} concentration. The parameters $g_{i,0}$ and $b_i$ (see Eq.~\ref{eq:FreeEnergyVersusSalt}) used in our numerical calculations were obtained from Sun \textit{et al.}~\cite{Sun2018}, who reported equilibrated concentrations of $T=3$, $T=4$, and free dimers at different CaCl\textsubscript{2} concentrations. From these equilibrium constants, the corresponding dimer binding energies $g_3$ and $g_4$ were calculated~\cite{Katen2013}, providing the salt-dependent inputs that drive our theoretical predictions.

The solid curves in Fig.~\ref{fig:fig_salt-energy} show the predictions of the kinetic model, which align closely with the experimental data across the full concentration range. By contrast, the dashed curves represent the thermodynamic model, which fails to capture the observed transition unless the salt-sensitivity parameter is assigned unrealistic values.
These results demonstrate that the kinetic model explains the experimental observations more accurately than the thermodynamic model and that quantitatively reproduces the salt-dependent shift in capsid morphology.

\subsection{On the Irreversibility of Salt-Driven Capsid Assembly}

\begin{figure}
\centering
\includegraphics[width=\linewidth]{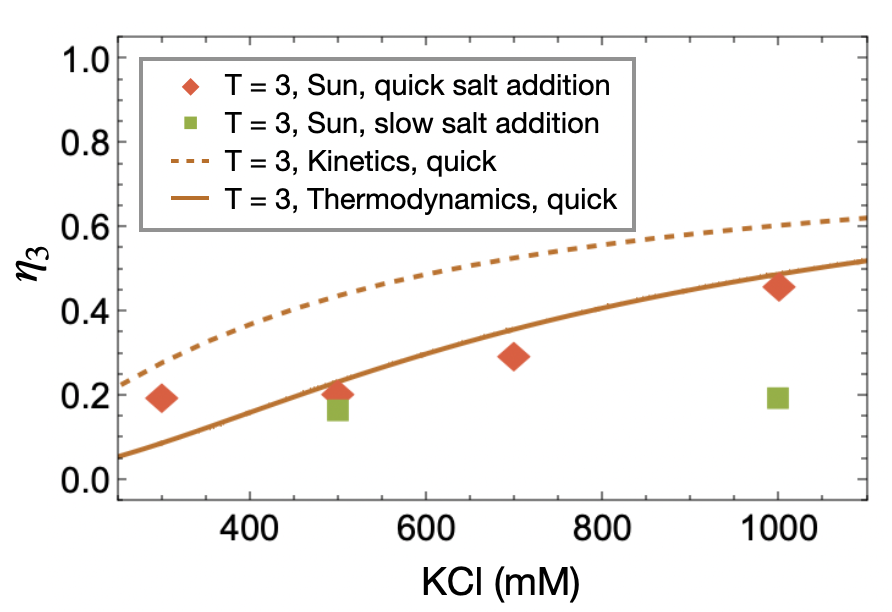}
\caption{ \textbf{The rate of salt addition influences the outcome of capsid assembly. } 
When salt is added rapidly (all at once at the beginning of the reaction), the relative $T=3$ population ($\eta_3$) increases with salt concentration, as shown by the red diamonds. In contrast, slow salt addition (100 mM increments every 12 hours until the target concentration is reached) produces little change in $\eta_3$, as indicated by the green squares. The dashed and solid curves represent kinetic and thermodynamic model predictions for rapid addition, respectively, both of which describe the experimental trend. The insensitivity of $\eta_3$ to gradual salt addition highlights the role of kinetic effects in capsid assembly. 
The fitting parameters used here are $g_3=-26.025+32.485/\sqrt{\mathrm{[KCl]}}$ and $g_4=-26.000+31.743/\sqrt{\mathrm{[KCl]}}$. 
\label{fig:fig_saltadditionrate}}
\end{figure}

A key implication of the thermodynamic model is the assumption that subunits can freely exchange between completed capsids and free dimers in solution. If this assumption holds, then altering the assembly environment -- such as by adding salt during the assembly process --should shift the relative abundance of $T=3$ and $T=4$ capsids as the system equilibrates. 

To examine this effect, Sun \textit{et al.}~\cite{Sun-Ionic2018} prepared two sets of salt-free samples to test whether changing the salt concentration during assembly would alter the relative abundance of $T=3$ capsids. In the first set, KCl was added all at once at different concentrations (rapid addition), and $\eta_3$ was measured. In the second set, KCl was added in 0.1 M portions at 12-hour intervals until the desired concentration was reached (gradual addition). The concentrations of $T=3$ and $T=4$ capsids are plotted in Fig.~\ref{fig:fig_saltadditionrate}. In the rapid-addition experiments, which began from free protein subunits, $\eta_3$ increased with salt concentration, whereas under gradual addition $\eta_3$ remained nearly unchanged. If subunits in completed capsids could freely exchange, then gradual salt addition would have allowed the system to adjust continuously toward the $T=3$-favoring equilibrium. However, the data in Fig.~\ref{fig:fig_saltadditionrate} show that slow salt addition does not alter the capsid structure once assembly is complete. This suggests that capsid assembly under these conditions is effectively irreversible, with structures determined at early stages of assembly and insensitive to free-energy changes introduced later. These findings indicate that once $T=3$ and $T=4$ capsids are formed, they remain kinetically trapped and stable despite changes in salt concentration.

Interestingly, the thermodynamic prediction (solid curve in Fig.~\ref{fig:fig_saltadditionrate}) aligns well with the experimental data for rapid salt addition. In this case, the screening parameters $b_3$ and $b_4$ used in Eq.\ref{eq:FreeEnergyVersusSalt} were approximated based on earlier fits to CaCl\textsubscript{2} data. By calculating the ionic strength with Eq.\ref{eq:ionicstrength}, we find that the Debye length in KCl solution is $\sqrt{3}$ longer than in CaCl\textsubscript{2} at the same molar concentration, \textcolor{black}{as expected}. The agreement with thermodynamic predictions suggests that the system equilibrates during assembly, but once capsids are complete, they can no longer dissociate even if salt conditions change.

\section*{Conclusion}
In this paper, we developed both thermodynamic and kinetic formalisms to investigate the self-assembly of HBV capsids, with particular emphasis on how salt concentration and dimer concentration influence the distribution of $T=3$ and $T=4$ morphologies. Our results demonstrate that the two approaches capture complementary aspects of the assembly process: the thermodynamic model successfully describes equilibrium distributions at intermediate dimer concentrations, while the kinetic model captures assembly dynamics, particularly at low concentrations.

The thermodynamic model accurately reproduces the equilibrium abundance of $T=3$ and $T=4$ capsids, including their sensitivity to electrostatic screening. Using Debye–Hückel theory, it captures how ionic strength modulates protein–protein interactions and explains experimental variations observed with both monovalent (KCl) and divalent (CaCl\textsubscript{2}) salts. However, the model breaks down below $c^*$ and cannot account for the fact that, once capsids are assembled, the energy barrier for dissociation is too high for them to disassemble and reassemble in response to changing salt conditions. These limitations highlight the need for a complementary approach.

Our kinetic model addresses these gaps by incorporating non-equilibrium effects. It not only explains the observed insensitivity of capsid distributions to salt concentrations after assembly, consistent with limited subunit exchange, but also captures the time-dependent trajectories of $T=3$ and $T=4$ formation across different protein concentrations. 

Together, these results demonstrate that integrating thermodynamic and kinetic perspectives is essential for a comprehensive understanding of HBV capsid polymorphism. This combined theory provides a more complete picture of symmetry selection and pathway accessibility, insights that may extend broadly to other self-assembling biological and synthetic systems.

A deeper understanding of the physical principles governing virus assembly and polymorphism will advance the rational design of self-assembling protein architectures with applications in virology, targeted drug delivery, and nanomaterial engineering. By bridging equilibrium theory and non-equilibrium dynamics, this work provides a foundation for understanding and ultimately controlling assembly in complex environments.

\section*{Supplementary Material}
The supplementary material contains additional information on the resistive-pulse sensing (RPS) analysis and the computational procedure used to derive dimer binding free energies from equilibrium constants for capsid formation.

\section*{Author Contributions}
Conceptualization, R.Z. and P.v.d.S.; Methodology, K.Y., J.M.G., A.R., R.Z. and P.v.d.S.; Formal Analysis, K.Y., J.M.G., A.R., R.Z. and P.v.d.S.; Investigation, K.Y., J.M.G., A.R., R.Z. and P.v.d.S.;  Data Curation, K.Y., J.M.G., A.R., R.Z. and P.v.d.S.; Writing – Original Draft Preparation, K.Y., R.Z. and P.v.d.S.; Writing – Review \& Editing, K.Y., R.Z. and P.v.d.S.; Visualization, K.Y., J.M.G., A.R., R.Z. and P.v.d.S.; Supervision, R.Z.; Project Administration, R.Z.; Funding Acquisition, R.Z.

\begin{acknowledgments}
This material is based upon work supported by the National Science Foundation under Grant No. NSF DMR-2131963 and Grant No. MCB/PHY-2413062. 
PvdS received support from the Institute of Complex Molecular Systems of Eindhoven University of Technology during the period that the work was done.
\end{acknowledgments}

\bibliography{article}

\clearpage

\end{document}